\documentstyle[multicol,epsf,aps]{revtex}
\begin{document}
\title{Maxwell Model of Traffic Flows}
\author{E.~Ben-Naim$\dag$ and P.~L.~Krapivsky$\ddag$}
\address{$\dag$Theoretical Division and Center for Nonlinear Studies, 
Los Alamos National Laboratory, Los Alamos, NM 87545, USA}
\address{$\ddag$Center for Polymer Studies and Department of Physics,
Boston University, Boston, MA 02215, USA}
\maketitle

\begin{abstract} 
  We investigate traffic flows using the kinetic Boltzmann equations
  with a Maxwell collision integral. This approach allows analytical
  determination of the transient behavior and the size distributions.
  The relaxation of the car and cluster velocity distributions towards
  steady state is characterized by a wide range of velocity dependent
  relaxation scales, $R^{1/2}<\tau(v)<R$, with $R$ the ratio of the
  passing and the collision rates. Furthermore, these relaxation time
  scales decrease with the velocity, with the smallest scale
  corresponding to the decay of the overall density. The steady state
  cluster size distribution follows an unusual scaling form $P_m \sim
  \langle m\rangle^{-4} \Psi(m/\langle m\rangle^2)$. This distribution
  is primarily algebraic, $P_m\sim m^{-3/2}$, for $m\ll \langle
  m\rangle^2$, and is exponential otherwise.
    
\end{abstract}

\begin{multicols}{2} 

\section{Introduction}

Traffic flows exhibit a variety of collective behaviors typical to
non-equilibrium systems\cite{Prigogine,hab,Leutz,wolf,dirk}.  The observed
phenomenology is rich and includes shock waves, traffic jams, clustering, and
synchronized flow\cite{wolf,dirk,hydr,kr}. A number of models and theoretical
approaches including fluid mechanics\cite{hab,hydr,kerner}, cellular
automata\cite{Nagel,Biham,Nagatani,SS,Nagel1,Nagel2,Brankov,Evans,Ktitarev},
particle hopping\cite{Benjamini,krug,Evans2}, ballistic
motion\cite{bal1,bal2,Pav,eps,ep,epnew,Nagatani1,Helbing}, and optimal
velocity\cite{opt1,opt2,jap} are used to describe these observations. Yet,
different approaches have different virtues, e.g., kinetic theory is more
appropriate for dilute flows, while fluid mechanics is more appropriate for
dense flows.

Here, we focus on the kinetic description of traffic. Previously, we
introduced a microscopic ballistic motion model and used it to derive
Boltzmann Equations (BE) for traffic on a one lane roadway\cite{eps}.
A generalization to situations where passing is allowed shows that a
transition from a low-density ``laminar'' flow to a high-density
``congested'' flow generally occurs\cite{ep,epnew}.  
The resulting BE are technically difficult, and a
number of important questions remain unresolved including the
transient characteristics and the cluster-size distribution.  Indeed,
previous studies addressed only steady state properties and the results
concerned mainly the velocity distributions.

Our goal is to obtain these relevant flow characteristics.  To this
end, we propose an approach inspired by Maxwell's classical model,
widely used in kinetic theory\cite{r,ernst}. This Maxwell approach
uses a velocity independent collision rate, thereby considerably
simplifying the analysis.  In fact, upon transforming the kinetic equations
from integral into differential ones, the Maxwell model results in
{\em first} order differential equations while the Boltzmann approach
leads to {\em second} order equations.

We will show that the Maxwell approximation is faithful to the nature
of the original traffic equations as it qualitatively reproduces
transient characteristics for no passing zones, as well as steady
state characteristics for passing zones. We further find that the
cluster velocity distribution approaches its steady state according to
a wide spectrum of relaxation scales, with the smallest describing
decay of the overall cluster density.  Furthermore, the size
distribution is characterized by a strong algebraic tail for small and
average sizes, while it is exponentially small for large size.

\section{The Maxwell Model}

The ballistic motion approach models the basic processes underlying
one lane traffic flows: passing and slowing down due to clustering.
The main assumption is that each driver has a fixed intrinsic
velocity.  The driving rules are as follows: A car moves with constant
intrinsic velocity on a one lane road until it overtakes a smaller
velocity car or a cluster.  After such an encounter, or ``collision'',
the incident car immediately adopts this smaller velocity, thereby
joining a cluster. Cars may also resume driving with their intrinsic
velocities, and such passing events are assumed to occur with a constant
rate. This model is an idealized description of one lane traffic flows
as several time and length scales including the actual collision time,
the passing time, and the car size are neglected.

Let $P(v,t)$ be the density of clusters moving with velocity $v$ at
time $t$, and let $P_0(v)$ be the intrinsic velocity distribution.
Natural initial conditions where cars are randomly distributed in
space and drive with their intrinsic velocities are imposed, i.e.,
$P(v,0)=P_0(v)$. Under the assumption that space and velocity remain
uncorrelated, a mean-field Boltzmann equation is written
\begin{eqnarray}
\label{BE}
{\partial P(v,t)\over \partial t}=&&t_0^{-1}\left[P_0(v)-P(v,t)\right]\\
-&&P(v,t)\int_0^v dv' U(v,v')P(v',t).\nonumber
\end{eqnarray}
The first term on the right-hand side represents cars escaping their
respective clusters with a constant rate $t_0^{-1}$. The last term
accounts for decrease in the cluster density due to collisions. For
traffic flows the collision rate should read $U(v,v')=v-v'$. For such
a collision rate, steady state properties of the velocity
distributions can be obtained by transforming the rate equation into a
second order nonlinear differential equation\cite{ep}. However, a
number of important characteristics including the size distribution
and time dependent properties appear to be analytically intractable.

We propose using a constant collision rate, $U(v,v')=u_0$ to simplify the
above traffic equations. Similar in spirit approximations, termed the Maxwell
Model (MM), proved useful in kinetic theory\cite{r,ernst}.  A natural choice
for the constant rate $u_0$ is the average velocity difference, $u_0=\langle
v-v'\rangle\propto\langle v\rangle$.  One may wonder whether such an
approximation is reasonable for traffic flows.  Ignoring $P(v',t)$ in the
collision integral $I(v)=\int_0^v dv' U(v,v')P(v',t)$, we have $I(v)\propto
v$ for the MM, while $I(v)\propto v^2$ for the BE.  Hence, the integral
remains an increasing function of the velocity. Furthermore, cars must slow
down before a collision, and therefore, the collision rate should be slower
than linear.  The MM can actually be considered as the limiting case of zero
deceleration, while the linear rate corresponds to the limit of infinite
deceleration.

Let $c_0$ be the initial car concentration, $v_0$ the average intrinsic
velocity, $t_0^{-1}$ the passing rate, and $u_0$ the collision rate.
Introducing the dimensionless velocity $v/v_0\to v$, space $xc_0\to x$, and
time $c_0v_0t \to t$ variables normalizes the initial concentration and
typical velocity to unity. The master equation (\ref{BE}) is characterized by
two dimensionless numbers
\begin{equation}
\label{ME}
{1\over \nu}
{\partial P(v,t)\over \partial t}\!=\!{1\over R}
[P_0(v)\!\!-\!\!P(v,t)]
-\!\!P(v,t)\!\int_0^v\!\!\!dv' P(v',t).
\end{equation}
The normalized collision rate, $\nu=u_0/v_0$, merely rescales time.
Thus, it is set to unity without loss of generality.  The number
$R=c_0u_0t_0=t_{\rm esc}/t_{\rm col}$, equals the ratio of the two
elementary time scales, the escape time $t_{\rm esc}=t_0$, and the
collision time $t_{\rm col}=(c_0u_0)^{-1}$.  This number, termed the
``collision number'', plays an important role in determining the
nature of traffic flows.

We will show below that the Maxwell model is completely solvable.
Although quantitative results of the MM may differ from the BE, they
faithfully reproduce the qualitative behavior of the traffic
equations.

\section{The Cluster Velocity Distribution}

We start with steady-state and time dependent properties of the
cluster velocity distribution.  Let us introduce the auxiliary
function
\begin{equation}
\label{Q}
Q(v,t)=R^{-1}+\int_0^v dv'P(v',t),
\end{equation}
which gives the cluster distribution via differentiation
$P(v,t)=\partial Q(v,t)/\partial v$.  This auxiliary function enables
us to reduce the integro-differential Eq.~(\ref{ME}) into the
differential equation
\begin{equation}
\label{Qvt1}
{\partial \over\partial t}{\partial Q\over\partial v}=
{1\over R}\,{\partial Q_0\over\partial v}-Q\,{\partial Q\over\partial v}.
\end{equation}
This equation can be integrated over $v$, and  using 
the boundary condition $Q|_{v=0}=R^{-1}$ we find 
\begin{equation}
\label{Qvt2}
{\partial Q(v,t)\over\partial t}=
{Q_0(v)\over R}-{Q^2(v,t)\over 2}-{1\over 2R^2}.
\end{equation}
Integrating Eq.~(\ref{Qvt2}), the auxiliary function is obtained explicitly
for arbitrary initial conditions
\begin{equation}
\label{Qsol}
Q(v,t)=Q_{\infty}(v){1+A(v)e^{-tQ_{\infty}(v)}\over 1-A(v)e^{-tQ_{\infty}(v)}},
\end{equation}
with notation $A(v)=[Q_0(v)-Q_{\infty}(v)]/[Q_0(v)+Q_{\infty}(v)]$.  Here we
use the subscript $\infty$ to denote steady state. The steady state auxiliary
function $Q_{\infty}(v)\equiv Q(v,t=\infty)$ is given by
\begin{equation}
\label{qsol}
Q_{\infty}(v)=R^{-1}\left[1+2R\int_0^v dv' P_0(v')\right]^{1/2}.
\end{equation}
Since the concentration is obtained from $Q(v,t)$ using
$c(t)=\lim_{v\to\infty}[Q(v,t)-R^{-1}]$, and since the cluster
velocity distribution is obtained by differentiation, Eq.~(\ref{Qsol})
represents a complete explicit solution of the Maxwell model.

We first examine steady state properties of the cluster velocity
distribution.  Comparing with the corresponding behavior emerging from
the BE will allow us to test the utility of the Maxwell model.
Evaluating the infinite velocity limit of the auxiliary function gives
the overall cluster density
\begin{equation}
\label{ceq}
c_{\infty}=R^{-1}\left(\sqrt{1+2R}-1\right).
\end{equation}
A remarkable feature of the steady state cluster density is that it is
a function of the collision number only. Such independence of the
initial velocity distribution has been observed in a few other
ballistic aggregation problems\cite{opt1,jp}.  Note that
$c_{\infty}=1-R/2+{\cal O}(R^2)$ for $R\ll 1$, i.e., the difference
from the initial density is of order $R$ in the laminar flow
regime. In this study, we will focus on the complementary nontrivial
limit of congested flows, i.e., $R\gg 1$. Here, the cluster
concentration is significantly reduced, $c_{\infty}\sim R^{-1/2}$, and
large clusters with an average size $\langle m\rangle
=c_{\infty}^{-1}\sim R^{1/2}$ form in agreement with the BE results.

The cluster velocity distribution is obtained from 
Eq.~(\ref{qsol}) by differentiation 
\begin{equation}
\label{pv}
P_{\infty}(v)=P_0(v)\left[1+2R\int_0^v dv' P_0(v')\right]^{-1/2}. 
\end{equation}
When $R\ll 1$, the difference between the initial and the steady state
distributions is of order $R$. This indicates a laminar flow regime when the
correction due to collisions is small and can be obtained by expanding the
solution perturbatively around the initial state. When $R\gg 1$, we use the
notation $I_0(v)=\int_0^v dv'P_0(v')$ and write the leading behavior of
Eq.~(\ref{pv}) as
\begin{equation}
\label{leading}
P_{\infty}(v)\simeq \cases{P_0(v)       &$v\ll v^*$;\cr
               P_0(v)[2RI_0(v)]^{-1/2}  &$v\gg v^*$.}
\end{equation}
The two limiting behaviors match at the threshold velocity $v^*$ which is
found from $1\sim RI_0(v^*)=R\int_0^{v^*} dv\,P_0(v)$.  In agreement with the
Boltzmann approach\cite{ep,epnew}, a boundary layer structure is found for
the velocity distribution, where in the inner region the original
distribution prevails, while in the outer region, the distribution is
substantially reduced.  The average cluster velocity remains of order unity.
Additionally, the suppression of the fastest velocities is proportional to
the concentration, again in agreement with the BE results.  We conclude that
although the MM differs quantitatively from the exact BE behavior, it
qualitatively reproduces the steady state behavior.

We turn now to analyzing the transient properties and in
particular the approach towards steady state.  The time dependent
concentration reads
\begin{equation}
\label{c}
c(t)=c_{\infty}{1+Be^{-t/\tau_c}\over 1-Be^{-t/\tau_c}},
\end{equation}
with the constant $B=A(\infty)=(1-c_{\infty})/(1+c_{\infty})$ and the
relaxation time $\tau_c=R/\sqrt{2R+1}$ corresponding to the concentration
decay.  We see that the cluster concentration exponentially approaches its
steady state value
\begin{equation}
\label{casymp}
c(t)\simeq c_{\infty}\left(1+2Be^{-t/\tau_c}\right).  
\end{equation}
As the distribution changes slightly in the laminar phase, relaxation
times remain of order unity when $R\ll 1$. However,
for congested flows the relaxation time diverges
with the collision number $\tau_c\sim R^{1/2}$. 

The explicit time dependent auxiliary function allows determination of
relaxation properties of the cluster velocity distribution. In the
long time limit Eq.~(\ref{Qsol}) reads
\begin{equation}
\label{Qrel}
Q(v,t)=Q_{\infty}(v)\left[1+2A(v)e^{-t/\tau(v)}\right]
\end{equation}
with the velocity dependent relaxation time $\tau(v)=1/Q_{\infty}(v)$.
Thus, the steady state properties are reflected in the transient
characteristics. The velocity dependence of the relaxation time
$\tau(v)$ becomes especially pronounced for large collision numbers
where it exhibits the following boundary layer structure
\begin{equation}
\label{tauv}
\tau(v)\sim\cases{R                &$v\ll v^*$;\cr 
                  [R/I_0(v)]^{1/2} &$v\gg v^*$.}
\end{equation}
For sufficiently small velocities, the collision integral is
negligible, and the relaxation time $R$, suggested by Eq.~(\ref{ME})
holds. While small velocities are governed by (almost) velocity
independent relaxation scales, large velocities are characterized by
velocity dependent decay rates. Furthermore, a large range of
relaxation scales exists $R^{1/2}<\tau< R$ with the larger relaxation
scales corresponding to smaller velocities. This is consistent with
dimensional arguments that time and velocity are inversely related.
Interestingly, the smallest possible relaxation scale corresponds to
the overall cluster density.

One anticipates that the relaxation time $\tau(v)$ also governs the decay of
the cluster velocity distribution $P(v,t)$. This is indeed the case.  To
obtain explicit expressions we first simplify Eq.~(\ref{Qrel}),  
\begin{equation}
\label{Qrel1}
Q(v,t)-Q_{\infty}(v)\simeq 
\cases{RI_0^2(v)e^{-t/\tau(v)}  &$v\ll v^*$;\cr 
2Q_{\infty}(v)e^{-t/\tau(v)}    &$v\gg v^*$.}
\end{equation}
Differentiating with respect to $v$ gives the cluster velocity 
\begin{eqnarray*}
P(v,t)\!-\!P_{\infty}\simeq \cases{
2RP_0(v)I_0(v)e^{-t/\tau(v)}          &$v\ll v^{**}$;\cr 
-RP_0(v)I_0^2(v)[te^{-t/\tau(v)}]     &$v^{**}\!\!\! \ll\!\! v\!\ll v^*$;\cr 
-2R^{-1}P_0(v)[te^{-t/\tau(v)}] &$v\gg v^*$;}
\end{eqnarray*}
with $\tau(v)$ given by Eq.~(\ref{tauv}).  The expressions match at the
boundary velocities which are determined from $RI_0(v^*)\sim 1$ and
$tI_0(v^{**})\sim 1$.  Only for velocities slower the decaying 
 boundary velocity $v^{**}(t)$ is the correction to the cluster
density positive.  This is surprising since both the overall cluster
density and the auxiliary function exhibit positive corrections, as
one would naively expect since $P_0(v)>P_{\infty}(v)$.

Since the relaxation times diverge with increasing $R$, an
intermediate behavior should emerge in the time range $t\ll \sqrt{R}$.
In this regime, the system has not yet ``realized'' that passing is
allowed, and the behavior should agree with the no passing case where
$R=\infty$. By directly solving Eq.~(\ref{Qvt2}) with $R^{-1}=0$ one
finds
\begin{equation}
\label{pvtinf}
P(v,t)={P_0(v)\over \left[1+{1\over 2}tI_0(v)\right]^2}.
\end{equation}
For arbitrary intrinsic velocity distribution, a scaling asymptotic
behavior emerges 
\begin{equation}
\label{Fvt}
P(v,t)\simeq {c(t)\over \langle v(t)\rangle}
F\left({v\over\langle v(t)\rangle}\right), 
\end{equation} 
with the cluster concentration $c(t)\sim t^{-1}$ and the average
velocity determined by $tI_0\left(\langle v(t)\rangle\right)\sim
1$. We see that the average velocity in the no passing case is
proportional to the time dependent boundary velocity: $\langle
v(t)\rangle\sim v^{**}$.  When the leading small velocity behavior of
the intrinsic velocity distribution is algebraic, $P_0(v)\sim v^{\mu}$
as $v\to 0$, the average velocity decays as a power law in time,
$\langle v(t)\rangle \sim t^{-\beta}$ with $\beta=1/(\mu+1)$.
Comparing with the exact behavior in the no passing limit of ballistic
motion model, we see that the overall scaling picture is reproduced,
while the quantitative details and in particular the scaling exponents
are different.

To summarize, explicit expressions for all cluster properties are possible
in the realm of the MM. The relaxation towards steady state occurs in
two stages.  The early one, $t\ll \sqrt{R}$, corresponds to a no
passing intermediate asymptotics. Then, the passing mechanism comes
into play, and the system approaches steady state.  This latter
relaxation is nontrivial in several ways. The decay is non-uniform in
time as a wide range of time scales are observed.  It is also
non-uniform in velocity as the cluster velocity distribution involves
three layers of greatly different width, i.e.  it exhibits the
so-called ``triple deck'' structure\cite{nayfeh}. The first layer
$v\ll v^{**}(t)$ (referred to as the lower deck) shrinks with time and
the velocity distribution in this deck approaches the steady state
exponentially from above.  In the middle and upper decks, the approach
towards steady state is from below and has a linear in time correction
to the exponential decay.

\section{The Car Velocity Distribution}

The cluster velocity distribution does not provide the observed
distribution of car velocities since all clusters -- large and small
-- are taken with equal weight. In what follows, we concentrate on the
car velocity distribution, which determines basic properties such as
the flux.

Within the framework of the MM, the car velocity distribution
satisfies
\begin{eqnarray}
\label{MEG}
{\partial G(v,t)\over \partial t}=&&R^{-1}[P_0(v)-G(v,t)]\\
-G(v,t)&&\int_0^v dv' P(v',t)+P(v,t)\int_v^{\infty}dv' G(v',t)\nonumber.
\end{eqnarray}
The escape term is the sum of a gain term $R^{-1}[P_0-P]$ and a loss term
$-R^{-1}[G-P]$.  In a collision between two clusters, all cars belonging to
the faster cluster acquire the slower cluster velocity.  Thus, in both
collision terms the argument of $P$ is smaller than the argument of $G$.  In
contrast with Eq.~(\ref{ME}) collisions can now lead to a gain in the car
velocity distribution.  Although the integration limits resemble those of the
previous kinetic equations \cite{Pav}, the collision terms are different, a
reflection of the different treatment of cars and clusters in this theory.
One can verify that Eq.~(\ref{MEG}) conserves the car density
$1=\int_0^{\infty} dv\,G(v,t)$. An alternative approach for obtaining
$G(v,t)$ involves the conditional velocity distribution $P(v,v',t)$. This
more detailed distribution can also be used to verify $G(v,t)$, and for
completeness, we detail its derivation in Appendix A.

Let us introduce the auxiliary function 
\begin{equation}
\label{defg}
g(v,t)=\int_v^{\infty} dv'\,G(v',t). 
\end{equation}
In terms of the auxiliary functions $g$, $Q$, and $Q_0$, Eq.~(\ref{MEG})
becomes
\begin{equation}
\label{gvt1}
{\partial \over \partial t}
{\partial g(v,t)\over \partial v}=-
{\partial\over \partial v}\left[{Q_0(v)\over R}+g(v,t)Q(v,t)\right].
\end{equation}
Integrating over the velocity and using $g_0(v)=Q_0(\infty)-Q_0(v)$
gives the master equation
\begin{equation}
\label{gvt2}
{\partial g(v,t)\over \partial t}=R^{-1}g_0(v)-g(v,t)Q(v,t).
\end{equation}

We first analyze the steady state properties which are obtained
immediately from Eq.~(\ref{gvt2})
\begin{equation}
\label{ginfty}
g_{\infty}(v)={g_0(v)\over RQ_{\infty}(v)}.
\end{equation}
Interestingly, this auxiliary function and the cluster velocity
distribution experience the same relative reduction at the steady
state, $g_{\infty}(v)/g_0(v)=P_{\infty}(v)/P_0(v)=1/RQ_{\infty}(v)$.
Differentiating $g_{\infty}(v)$, we get
\begin{equation}
\label{Gv}
G_\infty(v)=P_0(v)\,{1+R+R I_0(v)\over
\left[1+2RI_0(v)\right]^{3/2}}.
\end{equation}
In the congested phase, $R\gg 1$,  the car velocity
distribution has the following limiting behaviors:
\begin{equation}
G_\infty(v)\sim \cases{RP_0(v),              &$v\ll v^*$;\cr
                R^{-1/2}P_0(v)I_0^{-3/2}(v), &$v\gg v^*$.} 
\end{equation}
Thus, while the fast tail decay $R^{-1/2}$ agrees with the Boltzmann
equation approach\cite{ep}, the slow tail enhancement $R$ is larger
than the Boltzmann result where this enhancement is of the order
$R^{\alpha}$ with $0\leq\alpha\leq 1$.

The car velocity distribution immediately gives the average size of a
$v$-cluster
\begin{equation}
\label{mv}
\langle m(v)\rangle={1+R+RI_0(v)\over 1+2RI_0(v)},
\end{equation}
obtained from $\langle m(v)\rangle=G(v)/P(v)$.  As expected, the average
cluster size decreases with the velocity. The average cluster size obeys the
bounds $1\leq\langle m(v)\rangle\leq 1+R$, with the upper (lower) bound
achieved by the slowest (fastest) clusters. An additional quantity,
immediately derived from the car velocity distribution is the flux,
$J_{\infty}=\int dv\,v\,G_{\infty}(v)$. One can use Eq.~(\ref{Gv}) to find
\begin{equation}
\label{j}
J_\infty=\int_0^\infty dv\,{1-I_0(v)\over
\sqrt{1+2RI_0(v)}}.
\end{equation}
In the congested phase, the flux is proportional to the threshold velocity,
$J_\infty\sim v^*$, in agreement with the Boltzmann equation results.

We now focus on the time dependent behavior.  Integration of equation
(\ref{gvt2}) gives an explicit expression for $g(v,t)$ (for a
derivation, see Appendix B)
\begin{eqnarray}
\label{gvt}
{g(v,t)\over g_{\infty}(v)}={Q(v,t)\over Q_{\infty}(v)}
+{Q^2(v,t)-Q^2_{\infty}(v)\over Q_{\infty}(v)}
\left[{1\over I_0(v)}-{t\over 2}\right]
\end{eqnarray}
The relaxation of $g$ follows directly from the relaxation of $Q$
since $g(v,t)-g_\infty(v)\propto Q(v,t)-Q_{\infty}(v)$ when
$t\to\infty$. Using Eq.~(\ref{Qrel1}), we evaluate the leading
relaxation behavior of $g(v,t)$ 
\begin{eqnarray*}
g(v,t)\!-\!g_{\infty}\simeq \cases{
2Rg_0(v)I_0(v)e^{-t/\tau(v)}&$v\ll v^{**}$;\cr 
-Rg_0(v)I_0^2(v)[te^{-t/\tau(v)}]&$v^{**}\!\!\ll\!\! v\!\ll\! v^*$;\cr 
-2R^{-1}g_0(v)[te^{-t/\tau(v)}]&$v\gg v^*$.}
\end{eqnarray*}
Interestingly, the relaxation of the (properly normalized) cluster 
and auxiliary car velocity distribution are identical, 
$[g(v,t)-g_{\infty}]/g_0(v)\simeq [P(v,t)-P_{\infty}]/P_0(v)$.  
Relaxation of the car  velocity distribution is obtained from 
$G=-\partial g/\partial v$ 
\begin{eqnarray*}
{G(v,t)\over G_{\infty}(v)}-1\simeq \cases{
-2 e^{-t/\tau(v)}&$v\ll v^{**}$;\cr 
-I_0^2(v)[t^2e^{-t/\tau(v)}]&$v^{**}\!\!\ll\!\! v\!\ll\! v^*$;\cr 
-I_0(v)R^{-1}[t^2e^{-t/\tau(v)}]&$v\gg v^*$.}
\end{eqnarray*}

Thus an exponential relaxation with a velocity dependent time scale $\tau(v)$
underlies the approach of all velocity distributions towards steady state.
The car velocity distribution exhibits the triple deck structure similar to
that of the cluster velocity distribution. Some details change however; for
example, in the middle and upper decks the prefactor $t^2$ further slows down
the decay of $G(v,t)$.  The car velocity distribution approaches its steady
state always from below.

\section{The Size Distribution}

An important characteristic of traffic flows, the cluster size
distribution, has been determined analytically only in the no passing
limit \cite{eps}. We now address this issue in the framework of the
MM.  Let us consider $P_m(t)$ the cluster size distribution which 
evolves according to the following rate equation
\begin{eqnarray} 
\label{pmt} 
{\partial P_m(t)\over\partial t}
=&&R^{-1}[mP_{m+1}(v,t)-(m-1)P_m(t)]\\ 
+R^{-1}\delta_{m,1}&&[1-c(t)]-c(t)P_m(t)+{1\over 2} \sum_{i+j=m} P_i(t)P_j(t).\nonumber 
\end{eqnarray}
Terms proportional to $R^{-1}$ account for escape, while the rest
represent collisions.  The overall collision rate experienced by a
cluster, $c(t)$, is velocity-independent.  These rate equations are
reminiscent of an aggregation-fragmentation process\cite{bt,e}.
Indeed, collisions lead to cluster aggregation while passing events
split clusters.

Since aggregation and fragmentation are opposite mechanisms, their
combined effect generally leads to a steady state.  We leave the
ambitious task of a complete solution for the future, and restrict our
attention to the steady state where Eqs.~(\ref{pmt}) read
\begin{eqnarray}
\label{pmst}
c_\infty P_m&=&R^{-1}[mP_{m+1}-(m-1)P_m]\nonumber\\ 
               &+&R^{-1}\delta_{m,1}(1-c_{\infty})+
{1\over 2}\sum_{i+j=m}P_iP_j. 
\end{eqnarray}
It is useful to introduce the generating function
\begin{eqnarray} 
\label{fz}
{\cal F}(z)=c_{\infty}^{-1}\sum_m z^m P_m.
\end{eqnarray}
At the steady state, it satisfies the Riccati equation 
\begin{equation}
\label{ric1}
{\cal F}^2-2{\cal F}+z+{c_\infty\over 1-c_\infty}\,z(1-z)
{d\over dz}\left({{\cal F}\over z}\right)=0.
\end{equation}
The identity $(1-c_{\infty})/(Rc_{\infty}^2)=1$ was used in obtaining
this equation.  

Although we could not solve these equations generally,
most of the interesting features can be obtained by carefully
analyzing the leading terms in $R$.  The asymptotic relation
$c_\infty\simeq\sqrt{2/R}$ suggests that the last term 
in Eq.~(\ref{ric1}) is negligible.  Solving the
resulting quadratic equation subject to the boundary condition ${\cal
F}(1)=1$ gives ${\cal F}=1-\sqrt{1-z}$. Expanding this expression in
powers of $z$ we arrive at
\begin{equation}
\label{gamma}
P_m=c_\infty\,{\Gamma\left(m-{1\over 2}\right)
\over 2\Gamma\left({1\over 2}\right)\Gamma(m+1)},
\end{equation}
which simplifies to $P_m\simeq (2\pi R)^{-1/2}m^{-3/2}$ for $m\gg 1$.
However, this solution does not apply for very large $m$, or
equivalently near $z=1$. This follows e.g. from the conservation of
the car density, $\sum_m mP_m=1$, which implies that a crossover from
(\ref{gamma}) to the tail behavior should occur at the cutoff size
$m_c\sim \langle m\rangle ^2\sim R$.

To investigate the very large $m$ behavior  we have to return
to Eq.~(\ref{ric1}). Fortunately, in the proximity of $z=1$, i.e., when
$1-z\sim R^{-1}$, the generating function depends on a single scaling
variable 
\begin{equation}
\label{ric2}
1-{\cal F}=c_\infty\Phi(\zeta), \quad {\rm with}\quad
\zeta={1-z\over c_\infty^2}.
\end{equation}
This can be seen by balancing the leading terms in Eq.~(\ref{ric1}).
The scaling function $\Phi$ satisfies the Riccati equation
\begin{equation}
\label{ric3}
\zeta\,\Phi'(\zeta)=\zeta-\Phi^2
\end{equation}
subject to the boundary condition $\Phi(0)=0$.  Using the
transformation $\Phi(\zeta)=\phi(\zeta)/\phi'(\zeta)$ reduces
Eq.~(\ref{ric3}) to a second order linear differential equation
\begin{equation}
\label{ric5}
\zeta\,\phi''(\zeta)=\phi(\zeta).
\end{equation}
This is a solvable one-dimensional Shr\"odinger equation for a
particle with zero energy in a repulsive Coulomb potential.  Indeed, a
solution is found by reducing Eq.~(\ref{ric5}) to the Bessel
equation. Choosing the solution which satisfies the appropriate
boundary conditions, $\phi=0, \phi'(\zeta)=1$ at $\zeta=0$, one finds
$\phi(\zeta)=\sqrt{\zeta}\,I_1(2\sqrt{\zeta})$ with $I_1(x)$ the
modified Bessel function of order one.  Returning to $\Phi(\zeta)$ we
obtain
\begin{equation}
\label{Phi}
\Phi(\zeta)=2\zeta
\left[1+2\sqrt{\zeta}\,
{I_1'(2\sqrt{\zeta})\over I_1(2\sqrt{\zeta})}\right]^{-1}=\!\!
\sqrt{\zeta}\,{I_1(2\sqrt{\zeta})\over I_0(2\sqrt{\zeta})}.
\end{equation}
The last expression is derived using the
identities $I_1'(x)=I_0(x)-x^{-1}I_1(x)$ and
$I_0'(x)=I_1(x)$\cite{bo}.

The function $\Phi(\zeta)$ is the Laplace transform of the properly
scaled size distribution. Indeed, Eq.~(\ref{ric2}) implies 
$\sum P_m(1-z^m)=c_\infty^2\Phi[c_\infty^{-2}(1-z)]$
whose inversion yields the scaling form $P_m(R)=c_\infty^4\Psi(c_\infty^2
m)$.  Therefore, in the large $R$ limit the size distribution 
follows the scaling form 
\begin{equation}
\label{pmscl}
P_m \simeq {1\over \langle m\rangle^4}\Psi\left({m\over \langle m\rangle^2}
\right),
\end{equation}
with $\langle m\rangle=1/c_\infty\simeq \sqrt{R/2}$.  The scaling
function $\Psi(M)$ obeys $\Phi(\zeta)=\int_0^\infty
dM\,\Psi(M)\left[1-e^{-\zeta M}\right]$.  Differentiating both sides
with respect to $\zeta$ shows that $\Phi'(\zeta)$ is simply the Laplace
transform of $M\Psi(M)$
\begin{equation}
\label{lap}
\Phi'(\zeta)=\int_0^\infty dM\,M\Psi(M)\,e^{-\zeta M}.
\end{equation}
Consequently, the asymptotic behavior of the size distribution can
be determined from the corresponding asymptotics of $\Phi(\zeta)$. The
latter are found from Eq.~(\ref{Phi}): 
\begin{equation}
\label{Philimits}
\Phi(\zeta)\simeq\cases{\zeta^*(\zeta+\zeta^*)^{-1} &$\zeta\to -\zeta^*$;\cr
                        \sqrt{\zeta}                &$\zeta\to\infty$.\cr}
\end{equation}
The algebraic behavior of $\Phi(\zeta)$ at large $\zeta$ implies an
algebraic behavior of $\Psi(M)$ at small $M$; similarly, the pole at
$\zeta=-\zeta^*$ ($\zeta^*\cong 1.445796$\cite{pole}) implies an
exponential decay for large $M$:
\begin{equation}
\label{CMlimits}
\Psi(M)\simeq\cases{(4\pi)^{-1/2} M^{-3/2},  &$M\ll 1$;\cr
                    \zeta^*\exp(-\zeta^*M),  &$M\gg 1$.\cr}
\end{equation}
In terms of the original variables we have
\begin{equation}
\label{cmlimits}
P_m\simeq\cases{(2\pi R)^{-1/2}m^{-3/2},          &$m\ll R$;\cr
              4\zeta^*R^{-2}\exp(-2\zeta^*m/R),   &$m\gg R$.\cr}
\end{equation}
These two limiting behaviors match at $m\sim R$ where $P_m\sim
R^{-2}$.  Additionally, the value of the cutoff size, $m_c\sim R$,
agree with our previous findings.

In conclusion, the Maxwell equation approach allows explicit
calculations of the size distribution. It decays algebraically with
size for small and average clusters, and exponentially for very large
clusters. The interesting aspect of the size distribution concerns its
scaling form. If the typical and the average size would be the same, a
naive scaling argument $m/\langle m\rangle$ would underly the size
distribution. However, a different picture emerges where the scaling
variable is $m/\langle m\rangle^2$. Indeed, Eq.~(\ref{cmlimits}) is
consistent with a typical size of order unity, in contrast with the
growing average size $\langle m\rangle\sim \sqrt{R}$, a reflection of
the anomalous algebraic behavior of the size distribution below the
cutoff size.

\section{The Size-Velocity Distribution}

So far, we have addressed velocity and size distributions
separately. However, size and velocity are coupled in a nontrivial
manner, and for example, slower clusters should be larger than faster
ones.  We thus consider $P_m(v,t)$, the distribution of clusters of
size $m$ and velocity $v$. This joint distribution evolves according
to 
\begin{eqnarray} 
\label{pmvtM} 
{\partial P_m(v,t)\over\partial t}
&=&R^{-1}[mP_{m+1}(v,t)-(m-1)P_m(v,t)]\nonumber\\ 
&+&R^{-1}\delta_{m,1}[P_0(v)-P(v,t)]-c(t)P_m(v,t)\nonumber \\
&+&\int_v^{\infty} dv' \sum_{i+j=m} P_i(v',t)P_j(v,t). 
\end{eqnarray}
The car and cluster velocity distributions are simply the zeroth and
first moment of the size distribution, $P(v,t)=M_0(v,t)$ and
$G(v,t)=M_1(v,t)$, with $M_a(v,t)=\sum_m m^a P_m(v,t)$.  Consequently,
the respective evolution equations are recovered by summation of
Eqs.~(\ref{pmvtM}) over $m$. Furthermore, integration over the
velocities gives the size distribution and Eq.~(\ref{pmt}) is
recovered by using $P_m(t)=\sum_m P_m(v,t)$.

It proves useful to introduce the auxiliary functions $Q_m(v,t)=\int_v^\infty
dv'P_m(v',t)$.  The cluster size distribution can be expressed through these
auxiliary functions, $P_m(t)=Q_m(0,t)$. Additionally, the identity
$Q(v,t)+\sum Q_m(v,t)=R^{-1}+c(t)$ holds.  Integrating Eqs.~(\ref{pmvtM})
over $v$ gives
\begin{eqnarray}
\label{Qmvt}
{\partial Q_m(v,t)\over\partial t}
&=&R^{-1}[mQ_{m+1}(v,t)-(m-1)Q_m(v,t)]\nonumber\\ 
&+&R^{-1}\delta_{m,1}q(v,t)-c(t)Q_m(v,t)\\
&+&{1\over 2}\sum_{i+j=m}Q_i(v,t)Q_j(v,t)\nonumber
\end{eqnarray}
with $q(v,t)=\int_v^{\infty}dv'[P_0(v',t)-P(v',t)]$ or alternatively
$q(v,t)=1-c(t)+Q(v,t)-Q_0(v)$.  In deriving (\ref{Qmvt}) we used the
following boundary conditions: $Q_m=0$, $Q_0=1+R^{-1}$, and
$Q=c(t)+R^{-1}$ at $v=\infty$. Since the velocity plays the role of a
parameter, Eqs.~(\ref{Qmvt}) can be treated as ordinary differential
equations. We again restrict our attention to the steady state where
\begin{eqnarray}
\label{Qmsteady}
c_\infty Q_m(v)&=&R^{-1}[mQ_{m+1}(v)-(m-1)Q_m(v)]\nonumber\\ 
               &+&R^{-1}\delta_{m,1}q(v)+{1\over 2}\sum_{i+j=m}Q_i(v)Q_j(v),
\end{eqnarray}
with $q(v)=q_{\infty}(v)=1-c_{\infty}+Q_{\infty}(v)-Q_0(v)$.
Introducing the generating function
\begin{equation}
\label{gener}
{\cal Q}(z,v)=c_{\infty}^{-1}\sum_{m=1}^\infty z^m Q_m(v)
\end{equation}
reduces Eqs.~(\ref{Qmsteady}) into a set (parameterized by $v$) of
Riccati equations for ${\cal Q}={\cal Q}(z,v)$:
\begin{equation}
\label{ric}
{\cal Q}^2-2{\cal Q}+z{q(v)\over q(0)}
+{c_{\infty}\over 1-c_{\infty}}
{z(1-z)}{\partial \over\partial z}\left({{\cal Q}\over z}\right)=0
\end{equation}
This Riccati equation reduces to Eq.~(\ref{ric1}) when $v=0$. 
The above treatment of the size distribution suggests that
the derivative term in Eq.~(\ref{ric}) is negligible
for sufficiently small sizes.  In this case, Eq.~(\ref{ric})
simplifies to ${\cal Q}^2-2{\cal Q}+zq(v)/q(0)=0$ which is solved to
give ${\cal Q}(z,v)=1-\sqrt{1-zq(v)/q(0)}$.  Using the large $R$
behavior $q(v)/q(0)\to 1+Q(v)-Q_0(v)$ yields
\begin{equation}
\label{gQ3}
{\cal Q}(z,v)\simeq 1-\sqrt{1-z[1+Q(v)-Q_0(v)]}.
\end{equation}
Expanding  the expression on the right-hand side in powers of $z$
we arrive at 
\begin{equation}
\label{gQ4}
Q_m(v)\simeq P_m[1+Q(v)-Q_0(v)]^m,
\end{equation}
with $P_m$ the size distribution (\ref{gamma}).
The cluster size-velocity distribution is obtained by differentiating
the auxiliary distribution $Q_m(v)$ 
\begin{equation}
\label{pmv}
P_m(v)\simeq m P_m [P_0(v)\!-\!P(v)][1\!+\!Q(v)\!-\!Q_0(v)]^{m-1}
\end{equation}
Similar to the velocity distribution and the relaxation
scales, the size velocity distribution as well can be obtained
explicitly from the auxiliary function $Q(v)$. Consequently, it is
characterized by a boundary layer structure. The size-velocity
distribution is characterized by an exponential dependence upon the
size, with a velocity dependent prefactor. Additionally, there is an
algebraic prefactor that characterizes the overall size distribution.

The detailed analysis of the cluster size distribution suggests that
these results apply only for sufficiently small sizes.
Eqs.~(\ref{gQ3})-(\ref{pmv}) should hold as long as the (dropped) term
$R^{-1}[mQ_{m+1}(v,t)-(m-1)Q_m(v,t)]$ is negligible compared with the
(kept) term $c_\infty Q_m$. Using (\ref{gQ4}), the above approximation
is valid when
\begin{equation}
\label{condition}
m\ll \sqrt{R}\left[Q_0(v)-Q(v)\right].
\end{equation}
Hence, the range of validity of Eq.~(\ref{pmv}) strongly depends on the
cluster velocity. This can be seen using the average cluster size $\langle
m(v)\rangle=G(v)/P(v)=\sum_m m P_m(v)/\sum_m P_m(v)$, given by
Eq.~(\ref{mv}).  Estimating the same quantity from Eq.~(\ref{pmv}), gives the
correct leading large $R$ behavior when $v\gg v^*$, while it gives a
diverging average size rather than $\langle m(v)\rangle\sim R$ when $v\to 0$.
Indeed, the condition (\ref{condition}) is satisfied by the $\langle
m(v)\rangle$ only outside the boundary layer. Therefore the approximate
cluster size-velocity distribution is useful for small and average sizes when
$v\gg v^*$, while it holds only for sufficiently small sizes when $v\ll v^*$.
Obtaining the large size tail requires a more detailed analysis similar to
that performed for the size distribution.

\section{summary and outlook}

In summary, we introduced an approximation method for analyzing the
Boltzmann equations for one dimensional traffic flows. In analogy with
the Maxwell model (MM) of kinetic theory, we assumed a constant
collision rate. This approach results in first order (in the velocity)
differential equations.  Analysis of these equations leads to explicit
expressions for time dependent velocity distributions. Size-velocity
distributions can be determined in the steady state as well.  Although
there are some quantitative deviations, the overall qualitative
behavior including a boundary layer structure, existence of laminar
and congested phases, etc.  is in agreement with the results of the
original Boltzmann equations. Several quantities such as the size
growth exponent $1/2$ actually agrees with the Boltzmann equation. We
conclude that overall, the Maxwell approach is faithful to nature of
the problem and thus provides a useful approximation scheme.

The MM allows explicit calculation of several important features,
which are otherwise difficult to obtain. The approach towards the
steady state is generally exponential and is characterized by a wide
spectrum of velocity dependent relaxation scales, the smallest of
which corresponds to the overall cluster density.  The steady state
size distribution exhibits an unusual scaling form with a scaling
variable $m/\langle m\rangle^2$ rather than $m/\langle m\rangle$ which
is naively expected.  Additionally, the typical size which is of order
unity is much smaller than the average size which grows with the
collision number. This is a consequence of the algebraically diverging
distribution of small sizes.  This is an outcome of the
non-equilibrium nature of the steady state that does not satisfy
detailed balance as passing events reduce the cluster size by one
only, while clustering events can increase the cluster size by a large
number. This feature is independent of the details of the collision
mechanism and we expect the most features underlying the size
distribution to hold generally.

The MM can be refined and systematically improved.  Some of the quantitative
disagreements between the Maxwell and Boltzmann equation are rather obvious.
For example the correct value of the crossover velocity can be obtained by
replacing the integral $\int_0^v dv' P_0(v')$ with the integral $\int_0^v dv'
(v-v')P_0(v')$. This compensates for the approximate kernel taken in the MM
and results in the correct scaling exponents for the crossover velocity in
both passing and no passing zones.

Furthermore, an appropriate choice of the value of the prefactor $u_0$
reduces the discrepancies between the two approaches.  For example, the MM
gives a universal dependence of the density upon the collision number, $c\sim
R^{-1/2}$. However, for the BE if one assumes an algebraic intrinsic
distribution near the origin, $P_0(v)\sim v^\mu$ as $v\to 0$ different
behaviors are found for positive and negative $\mu$ \cite{ep}.  For $\mu>0$,
the density exhibits the universal behavior, $c_{\infty}\sim R_0^{-1/2}$,
while for $\mu<0$ the density becomes $\mu$ dependent, $c_{\infty}\sim
R_0^{-(\mu+1)/(\mu+2)}$. Here $R_0=c_0v_0t_0$ is the collision number within
the Boltzmann framework.  Choosing $u_0=\langle v\rangle=R_0^{\mu/(\mu+2)}$
(the actual BE behavior) implies $R=\nu R_0\sim R_0^{(2\mu+2)/(\mu+2)}$, and
thence $c_{\infty}\sim R_0^{-(\mu+1)/(\mu+2)}\sim R^{-1/2}$.  Therefore the
BE and MM results are consistent with each other if the appropriate choice
for the collision rate $u_0=\langle v\rangle$ is made.

Additionally, it would be interesting to compare the MM with the
actual traffic process.  Although the BE description is plausible at
the steady state, it is clearly an approximation for the transient
regime. For example, the BE differs from the exact behavior in the no
passing case.  Another avenue for further research is inhomogeneous
traffic flows where a hydrodynamic description may prove useful.  The
hydrodynamic framework should involve a multicomponent fluid
parametrized by the cluster size $m$. Specifically, the macroscopic
description requires the density $P_m(x,t)$, the average velocity
$u_m(x,t)=P^{-1}_m(x,t)\int dv\,vP_m(v,x,t)$, and the ``temperature''
(the average velocity square) for each $m$.  Such infinite-fluid
hydrodynamics may be quite different from the conventional one-fluid
hydrodynamics. Indeed, Eq.~(\ref{mv}) shows that the velocity
decreases as the mass increases. Similar results apply for the
temperature and thus, the equipartition of ``energy'' breaks down as
well, in contrast with usual hydrodynamics.

In conclusion, the MM is a useful approximation to the kinetic traffic
equations. This approach may be applicable to other traffic problems
as well.  In particular, it will be interesting to apply the Maxwell
approach to traffic models with more realistic passing mechanisms.

\vspace{.1in}
\noindent
The work of PLK has been supported by NSF (grant DMR-9632059) and ARO
(grant DAAH04-96-1-0114).

\end{multicols}

\appendix
\section{The conditional velocity distribution}

The car velocity distribution involves the leading car as well as the
rest of slowed down cars in the cluster. The former is described by
the cluster velocity distribution, while the latter is represented by
$P(v,v',t)$, the density of cars of intrinsic velocity $v$ driving
with velocity $v'$.  For the Maxwell model, the master equation for
this conditional distribution reads
\begin{equation}
\label{pvvt}
{\partial P(v,v',t)\over \partial t}
=-R^{-1}P(v,v',t)+P(v,t)P(v',t)
+P(v',t)\int_{v'}^v dv''P(v,v'',t)
-P(v,v',t)\int_0^{v'} dv'' P(v'',t).
\end{equation}
The first term accounts for loss due to escape, while the rest of the
terms represent changes due to collisions. Integrating these equations
over the first velocity index and using the relation
$G(v)=P(v)+\int_v^{\infty}\!\!dw\,P(w,v)$, one indeed recovers the
rate equation (\ref{MEG}).

Let us introduce the auxiliary function $Q(v,v',t)=\int_{v'}^v
dw\,P(v,w)$ which gives the conditional velocity distribution by
differentiation $P(v,v',t)=-\partial Q(v,v',t)/\partial v'$. This
auxiliary function evolves according to
\begin{equation}
\label{AQ1}
-{\partial\over \partial t}\,{\partial Q(v,v',t)\over \partial v'}=
Q(v',t)\,{\partial Q(v,v',t)\over \partial v'}
+P(v,t)\,{\partial Q(v',t)\over \partial v'}
+Q(v,v',t)\,{\partial Q(v',t)\over \partial v'}.
\end{equation}
Integrating Eq.~(\ref{AQ1}) over $v'$ and using the  
boundary condition $Q(v,v,t)=0$, we get
\begin{equation}
\label{AQ2}
-{\partial Q(v,v',t)\over \partial t}=
Q(v',t)Q(v,v',t)
+P(v,t)Q(v',t)-P(v,t)Q(v,t).
\end{equation}
This is a linear inhomogeneous differential equation for the auxiliary
function $Q(v,v',t)$ which includes already known cluster velocity
distributions.  Integrating Eq.~(\ref{AQ2}) we arrive at
\begin{equation}
\label{AQ3}
Q(v,v',t)=\int_0^t dt'\,P(v,t')[Q(v,t')-Q(v',t')]\,\,
\exp\left[-\int_{t'}^t dt''\,Q(v',t'')\right].
\end{equation}
The exact solution (\ref{AQ3}) can in principle be reduced to a more
explicit expression by following the procedure detailed in Appendix B
for transforming the formal solution of Eq.~(\ref{gvtf}) into
Eq.~(\ref{Agvt}).  Such a solution is very cumbersome so we do not
give it here.

The steady state conditional distribution is obtained immediately from
Eq.~(\ref{AQ2}) $Q(v,v')=P(v)[Q(v)/Q(v')-1]$.  The joint distribution
is found by differentiation, $P(v,v')=-\partial Q(v,v')/\partial
v'=P(v)P(v')Q(v)/Q^2(v')$, or explicitly
\begin{equation}
\label{Psolution}
P(v,v')={RP_0(v)P_0(v')\over\left[1+2RI_0(v')\right]^{3/2}}.
\end{equation}
In the laminar phase, this conditional distribution is proportional to
$R$, while it is algebraically suppressed in the congested phase.  One
can verify that Eq.~(\ref{Psolution}) is consistent with the cluster
and car distributions using the relations $P(v)=P_0(v)-\int_0^v dv'
P(v,v')$, and $G(v)=P(v)+\int_v^{\infty}\!\!dw\,P(w,v)$, respectively.

\section{The auxiliary car velocity distribution}

The master equation (\ref{gvt1}) for $g(v,t)$ can be integrated formally 
\begin{equation}
\label{gvtf}
g(v,t)=g_0(v)\Bigg[\exp\left(-\int_0^t dt'\,Q(v,t')\right)
+R^{-1}\int_0^t dt'\,
\exp\left(-\int_{t'}^t dt''\,Q(v,t'')\right)\Bigg].
\end{equation}
To obtain more explicit results, we notice that the velocity plays the
role of a parameter in Eq.~(\ref{gvtf}).  We thus change variable from
$t'$ to $q\equiv Q(v,t')$ and for example
\begin{eqnarray*}
\int_0^t dt'\,Q(v,t')=-2 \int_{Q_0}^Q dq\,
{q\over q^2-Q_{\infty}^2}=\ln {Q_0^2-Q_{\infty}^2\over Q^2-Q_{\infty}^2}.
\end{eqnarray*}
This change of variables allows us to perform the integration 
\begin{eqnarray}
\label{Agvt}
g(v,t)&=&g_0\left[{Q^2-Q_{\infty}^2\over Q_0^2-Q_{\infty}^2}-2
{Q^2-Q_{\infty}^2\over R}\int_{Q_0}^Q {dq\over (q^2-Q_{\infty}^2)^2}
\right]\nonumber\\
&=&g_0\left[{Q^2-Q_{\infty}^2\over Q_0^2-Q_{\infty}^2}+
{Q\over RQ_{\infty}^2}
-{Q_0\over RQ_{\infty}^2}{Q^2-Q_{\infty}^2\over Q_0^2-Q_{\infty}^2}
+{Q^2-Q_{\infty}^2\over 2RQ^3_{\infty}}\ln 
\left({Q-Q_{\infty}\over Q+Q_{\infty}}\Big/
{Q_0-Q_{\infty}\over Q_0+Q_{\infty}}\right)\right]\nonumber\\
&=&g_0\left[{Q\over RQ_{\infty}^2}+
{Q^2-Q_{\infty}^2\over I_0^2}
\left(1-{Q_0\over RQ_{\infty}^2}\right)
+{Q^2-Q_{\infty}^2\over 2RQ_{\infty}^3}
\ln (e^{-tQ_{\infty}})\right]\nonumber\\
&=&{g_{\infty}\over Q_{\infty}}
\left[Q+(Q^2-Q^2_{\infty})\left({1\over I_0}-{t\over 2}\right)\right].
\end{eqnarray}
In the above derivation we used the identities
$g_{\infty}=g_0/RQ_{\infty}$, $Q_0^2-Q_{\infty}^2=I_0^2$, and
$[1-Q_0/RQ_{\infty}^2]=I_0/RQ_{\infty}^2$.


\begin{thebibliography}{99}
\bibitem{pole} 
      The pole is related to the first zero $x_1\cong 2.40483$ 
      of the Bessel function $J_0(x)$ through $\zeta^*=x_1^2/4$.
\end{thebibliography}
\end{document}